\documentclass[reprint,aps,amsmath,amssymb,superscriptaddress]{revtex4-1}
\usepackage{graphicx}
\usepackage{dcolumn}
\usepackage{bm}
\usepackage{color,hyperref}

\newcommand{\fuzhou}{Department of Physics, Fuzhou University, Fuzhou 350108, Fujian, China }
\newcommand{\fujian}{Fujian Science and Technology Innovation Laboratory for Optoelectronic Information of China, Fuzhou 350108, Fujian, China}
\newcommand{\Italy}{Consiglio Nazionale delle Ricerche, Istituto dei Sistemi Complessi, via Madonna del Piano 10, I-50019 Sesto Fiorentino, Italy}
\newcommand{\italy}{Istituto Nazionale di Fisica Nucleare, Sezione di Firenze, via G. Sansone 1, I-50019 Sesto Fiorentino, Italy}

\begin{document}


\title{Heat conduction in a three-dimensional momentum-conserving
fluid}



\author{Rongxiang Luo}
\email[]{phyluorx@fzu.edu.cn}
\affiliation{\fuzhou}
\affiliation{\fujian}
\author{Lisheng Huang}
\affiliation{\fuzhou}
\affiliation{\fujian}
\author{Stefano Lepri}
\email[]{stefano.lepri@isc.cnr.it}
\affiliation{\Italy}
\affiliation{\italy}


\date{\today}

\begin{abstract}
Size-dependence of energy transport and the effects of reduced dimensionality on transport coefficients are of key importance for understanding nonequilibrium properties of matter on the nanoscale. Here, we perform nonequilibrium and equilibrium simulations of heat conduction in a 3D fluid with the multiparticle collision dynamics, interacting with two thermal-walls. We find that the bulk 3D momentum-conserving fluid has a finite non-diverging thermal conductivity. However, for large aspect-ratios of the simulation box,  a crossover from 3D to one-dimensional (1D) abnormal behavior of the thermal conductivity occurs. In this case, we demonstrate a transition from normal to abnormal transport by a suitable decomposition of the energy current. These results not only provide a direct verification of Fourier's law but also further confirm the validity of existing theories for 3D fluids. Moreover, they indicate that abnormal heat transport persists also for almost 1D fluids over a large range of sizes.
\end{abstract}


\maketitle



\textit{\textbf{Introduction.}} Understanding thermal transport is of key importance both for basic science and for technological development~\cite{S. Lepri2016,Benenti2020,N. B. Li2012,Z. W. Zhang2020,MacDonald1978}. Macroscopic behavior of thermal transport in the linear response regime is governed by Fourier's law
\begin{equation}\label{Eq1}
  J=-\kappa \nabla T,
\end{equation}
where $J$ is the heat current, $\nabla T$ is the spatial temperature gradient, and $\kappa$ is a finite constant termed as the thermal conductivity. The heat conduction behavior is also known as ``normal heat conduction" if it obeys the Fourier law or ``abnormal heat conduction" otherwise. \par

It is well established that for low-dimensional systems with a conserved total momentum, heat conduction is generically abnormal~\cite{S. Lepri2016}. Specifically, for one-dimensional (1D) momentum-conserving systems, $\kappa$ usually diverges with the system length $L$ as a power-law: $\kappa\sim L^{\alpha}$ \cite{S. Lepri1997}. The exponent $\alpha=1/3$ is related to the fact that fluctuations of the conserved field belong to the universality class of the famous Kardar-Parisi-Zhang (KPZ) equation
\cite{H. van Beijeren2012,Spohn2014}. Thus, the phenomenon should be largely independent of the microscopic details. For anharmonic chains this has been well numerically verified
in several works both for hard-core \cite{A. Dhar2001,G. Casati2003,Cipriani2005,S. Chen2014,Hurtado2016} and smooth interaction potentials ~\cite{Das2014,Wang2016}. Actually, a different $\alpha$ value signals a different universality class, as it may occur at specific points of the phase diagram or for systems with a different set of conserved quantities. A known instance is the case of nonlinear chains with even potentials, where the value of $\alpha\neq 1/3$ ~\cite{G. R. Lee-Dadswell2005,L. Delfini2006,Wang2011} although its precise value of $\alpha$ is not yet fully assessed theoretically (see Ref.\cite{Benenti2020} and references therein for a review).\par

For two-dimensional (2D) momentum-conserving systems, a $\kappa\sim \ln L$ divergence of the thermal conductivity is predicted from different theoretical apporaches~\cite{S. Lepri1998,O. Narayan2002} and also from an exactly solved stochastic model~\cite{G. Basile2006}. As for 1D case,
such prediction has been numerically verified in 2D lattices~\cite{A. Lippi2000,Delfini2005,L. Yang2006,D. Xiong2010}, and only recently, also in 2D fluids~\cite{P. D. Cintio2017,R. X. Luo2020}. However, there are some cases in which a power-law divergence have been reported, see~\cite{Wang2020} and references therein. Therefore, it is worth noting that these theoretical predictions equally apply to both lattices and fluids. With respect to the latter case, abnormal heat transport is related to the famous long-time-tail problem, long known in statistical mechanics \cite{Pomeau1975}.~\par

For the three-dimensional (3D) momentum-conserving systems, which is the subject of the present paper, all analytical predictions of existing theories~\cite{S. Lepri2016,G. Basile2006} support that the heat conduction behavior is normal. In 2010, the first numerical evidence for the validity of Fourier's law  was given by Saito and Dhar ~\cite{K. Saito2010}. They performed nonequilibrium simulations of heat conduction in a 3D anharmonic lattice and found that the system, under certain conditions, has a finite nondiverging $\kappa$ but can exhibit a crossover from 3D to 1D behavior. In the same year, similar results for the same class of lattice systems were obtained by Wang \emph{et al}. by means of equilibrium simulations~\cite{L. Wang2010}. However, to the best of our knowledge, at present only the 3D lattices have been examined. Thus, to verify whether the above picture of low-dimensional heat conduction applies equally well to both lattices and fluids, numerical evidences for the counterpart 3D fluids are still in urgent need. This then raises the relevant question: Can the finite thermal conductivity also be confirmed in 3D fluids? Also, can the dimensional-crossover behavior of heat conduction be observed in 3D fluids?\par

In this work, we give a positive answer to the above two questions.
To this aim, we consider a 3D system of particles in a cuboid-shaped box, interacting through the multi-particle collision (MPC) dynamics~\cite{A. Malevanets1999}. The MPC method
consists in replacing the conventional deterministic molecular dynamics with a stochastic rotation of velocities. It can correctly capture the hydrodynamic equations with
great advantages in the numerical simulations~\cite{J. T. Padding2006}. By using the MPC dynamics, researchers achieved a considerable understanding of various aspects of transport~\cite{G. Gompper2009}. More recently, it has been adopted to study the coupled particle and heat transport~\cite{Benenti2014,R.X. Luo2018}. Importantly, the MPC dynamics conserves the total momentum and energy of the system, and thus it can help us test the theoretical conjecture for momentum-conserving systems~\cite{P. D. Cintio2017,R. X. Luo2020}.\par

Here, we study heat transport properties of the 3D MPC fluid by means of extensive numerical simulations. We employ the nonequilibrium thermal-wall method where the system interacts with two heat baths at different temperatures and compare the results with those obtained by equilibrium Green-Kubo method. We show that the 3D fluid has finite thermal conductivity.  Upon increasing the aspect ratio of the simulation box, the fluid system can also exhibit the crossover from 3D to 1D behavior of the thermal conductivity.\par

\textit{\textbf{The 3D fluid model.}} We consider a 3D system of $N$ interacting point particles with equal mass $m$. All particles are confined in a cuboid-shaped volume of length $L$, width $W$, and height $H$ in the $x$, $y$ and $z$ coordinate, respectively.
At $x=0$ and $x=L$ in longitudinal direction, the particles exchange heat with two heat baths at temperatures $T_h$ and $T_c$, modeled as thermal walls ~\cite{Lebowitz1978}. When a particle hits the $x=0$ ($x=L$) boundary (of area $WH$), it is reflected back with new velocity components (denoted by $v_x$, $v_y$ and $v_z$ of its $x$, $y$ and $z$ directions, respectively) drawn from a distribution with probability densities~\cite{Lebowitz1978}
\begin{equation}\label{Eq2}
\begin{aligned}
  P\left(v_{x}\right)= & \frac{ m|v_{x}|}{k_{B}T_\iota}\textrm{exp}\left(-\frac{mv^{2}_{x}}{2k_{B}T_\iota}\right), \\
  P(v_{y,z})= &\sqrt{\frac{m}{2\pi k_{B}T_\iota}}\textrm{exp}\left(-\frac{mv^{2}_{y,z}}{2k_{B}T_\iota}\right),
\end{aligned}
\end{equation}
where $T_\iota$ ($\iota=h, c$) is the temperature of the respective heat bath in dimensionless units and $k_{B}$ is the Boltzmann constant. In the $y$ and $z$ directions the particles are subject to periodic boundary conditions (even though, it has been verified numerically that the results also apply to fixed boundary conditions, since in both cases $v_{y,z}>0$ and $v_{y,z}<0$ are with the same probability.).\par

The evolution in the bulk occurs by MPC dynamics~\cite{A. Malevanets1999,J. T. Padding2006,G. Gompper2009}, which simplifies the numerical simulation of interacting particles by coarse-graining the time and space at which interaction occur. Specifically, by MPC the system evolves in discrete time steps, consisting of non-interacting propagation during a time $\tau$ followed by instantaneous collision events. During the propagation, a particle keeps its velocity $\mathbf{v}_i$ unchanged and its position is
updated as $\mathbf{r}_i\rightarrow\mathbf{r}_i+\tau\mathbf{v}_i$.
As for the collisions, the system's volume is partitioned into cubic cells of linear size $a$ and for all particles in a cell, their velocities are rotated around a randomly chosen axis, with respect to their center of mass velocity $\mathbf{V}_{CM}$ by an angle, $\theta$ or $-\theta$, randomly chosen with equal probability. The velocity of a particle in a cell is thus updated from $\mathbf{v}_i$ to $\mathbf{V}_{CM}+\hat{\mathcal{R}}^{\pm\theta}\left(\mathbf{v}_i-\mathbf{V}_{CM}\right)$,
where $\hat{\mathcal{R}}^{\theta}$ is the rotation operator
by the angle $\theta$. Such moves preserve the total momentum and  energy of the fluid system.
The time interval between successive collisions $\tau$ and the collision angle $\theta$ tune the strength of the interactions and consequently affect the transport of the particles. Note that the angle $\theta=\pi/2$ corresponds to the most efficient mixing of the particle momenta. \par

In our simulations, we set $T_\iota$ to be slightly biased from the nominal temperature $T$, i.e., $T_{h,c}=T\pm\Delta T/2$, to measure the temperature profile $T(x)$, where $x$ is the space variable, and the heat current $J$. Numerically, $T(x)$ and $J$ are obtained in a same way as in the 2D case~\cite{R. X. Luo2020}. The thermal conductivity is finally computed by assuming Fourier's law, as $\kappa\approx JL/(T_h-T_c)$, since we have checked that in the linear response regime, the temperature jump between the heat bath and simulated system's boundary can be neglected (as shown in Fig.~\ref{fig3} below). We set the main parameters as follows: $m=T=k_B=1$, $\Delta T=0.2$, $a=0.1$, $\theta=\pi/2$, $\tau=0.1$, and the averaged particle number density $\rho=N/(LWH)=88$. Note that the average distance, $\lambda=\tau\sqrt{k_BT/m}$, that particles stream between rotations is equal to $a$, meaning that our simulations satisfy Galiean-invariance of the stochastic rotation dynamics~\cite{T. Ihle2001}. We set $W=H$ henceforth, and focus only on the the heat conduction in the $x$ direction for different $W$ and $L$. So, the relevant parameter is the cell aspect-ratio $W/L$. Relatively small values of it correspond to quasi-1D setups. In addition, long enough relaxation times $(>10^7)$ and integration times $(>10^8)$ are utilized to ensure the system reaches the steady state and the relative errors of all statistical averages are smaller than 0.5$\%$.\par

\begin{figure}
\includegraphics[width=10cm]{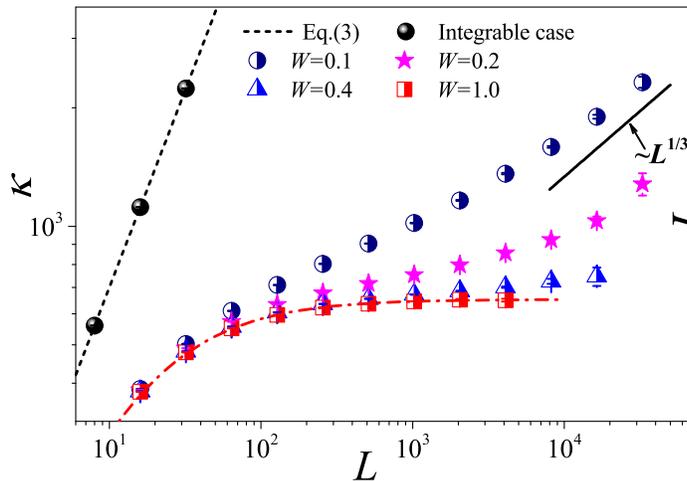}
\caption{(Color online) The thermal conductivity $\kappa$ as a function of the system length $L$ for the 3D fluid system with different $W$ values. The symbols are for the numerical results, and for reference the black solid line indicates the divergence with $L$ as $\sim L^{1/3}$. The red dot-dashed line is a fit of the data with $W=1$ with the functional expression $\kappa_{kin}=652.25/(1+12/L)$. }
\label{fig1}
\end{figure}

\textit{\textbf{Numerical results.}} First of all, we provide  evidence that in the ballistic (integrable) limit the thermal conductivity is a linear function of the system length. If all particles do not interact, i.e., each particle keeps its velocity unchanged from one heat bath to the other as it crosses the system, then the system is integrable. Under this situation,  $J$ can be
computed by a similar analysis as done in~\cite{R. X. Luo2019}. Taking it into Eq.~(\ref{Eq1}), we can obtain an analytical expression for the thermal conductivity:
\begin{equation}\label{Eq3}
\kappa=\rho L\sqrt{\frac{8k^{3}_B}{\pi m }}/\left(\frac{1}{\sqrt{T_h}}+\frac{1}{\sqrt{T_c}}\right).
\end{equation}
In Fig.~\ref{fig1} this result is compared with our simulations (black circles) and the agreement is perfect.\par


Now, let us turn to the interacting systems with collisions. It can be seen in Fig.~\ref{fig1} that for $W=1$, $\kappa$ tends to saturate and becomes constant as $L$ is increased. This result indicates that heat conduction for the 3D momentum-conserving fluid system is governed by the Fourier law, as expected~\cite{S. Lepri2016,G. Basile2006}. However, it can also be seen that as $W$ decreases, $\kappa$ is no longer constant but diverges with $L$. In particular, for $W=0.1$, $\kappa$ eventually approaches the scaling $\kappa\sim L^{1/3}$  predicted for 1D momentum-conserving fluids~\cite{S. Chen2014,O. Narayan2002}. Not surprisingly, the convergence is relatively slow and is affected by finite-size corrections, since anomalous scaling is expected to set in only at large scales. Note also that the scaling holds on a range of about two decades. We also remark that the dimensional crossover from small to large aspect ratio is similar to what observed in 3D lattices~\cite{K. Saito2010,L. Wang2010}.


\begin{figure}
\includegraphics[width=10cm]{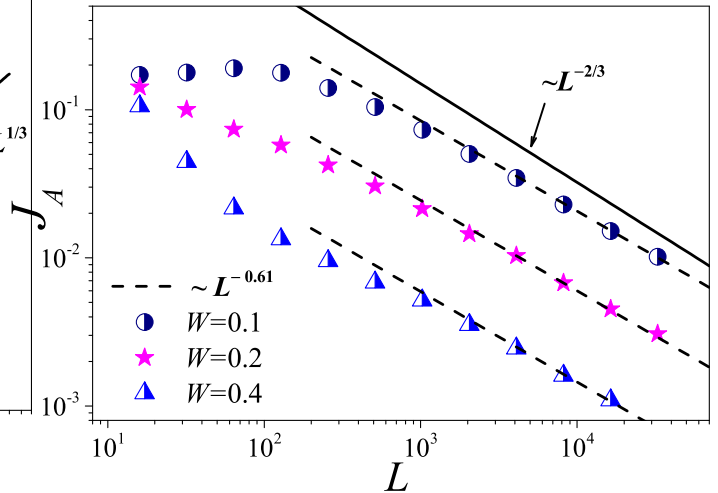}
\caption{(Color online) Scaling of the abnormal component of the heat currents $J_A$ for three $W$ values. $J_A$ is computed subtracting the normal component $J_N$ as detailed in the text. The dashed line is a best-fit of the data, yielding a decay $J_A\propto L^{-0.61}$,
while the solid line is the expected decay $L^{-2/3}$.}
\label{fig2}
\end{figure}

To further support the scenario of quasi-1D transport, we perform a similar analysis as done in~\cite{S. Lepri2020}. The idea is to decompose the heat current in a ``normal'' and ``abnormal'' components, i.e., $J=J_N+J_A$. To estimate the abnormal part, we first numerically compute the heat current $J$ and subtract from it the best estimate of the ``normal'' heat current, obtained as $J_N=\kappa_{kin}\Delta T/L$, where $\kappa_{kin}=\kappa_{0}/(1+A/L)$. Such formula is given by kinetic theory to account for the initial ballistic regime. The parameter $A$ is related to boundary resistance between the fluid
and the reservoirs~\cite{K. Aoki2001,S. Lepri2003}, and$\kappa_{0}$ is the
bulk value of the conductivity. As shown in Fig.\ref{fig1} the fit of the data with $W=1$ is very accurate meaning that the abnormal contribution to the thermal conductivity is negligible over the considered length range. The data in Fig.~\ref{fig2} show that for all small $W$ values, $J_A\sim L^{-0.61}$ when $L\rightarrow\infty$, approaching the expected scaling $L^{-2/3}$, the value predicted for 1D systems with the momentum conservation~\cite{S. Lepri2016}. We thus conclude that the measurements in 3D fluid system confirm the normal heat conduction and the crossover from 3D to 1D behavior of the thermal conductivity.\par

\begin{figure}
\includegraphics[width=10cm]{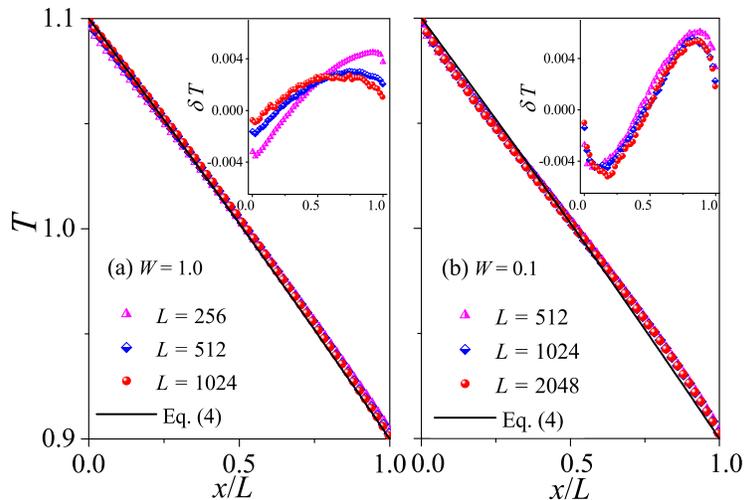}
\caption{(Color online) Plot of temperature profiles for the 3D fluid system with different $L$ values. Here our numerical results are compared with the analytical Eq.~(\ref{Eq4}). In (a) and (b) we fix $W=1.0$ and $W=0.1$, respectively. Inset: Plot of the differences $\delta T$ between the data and the black line.}
\label{fig3}
\end{figure}

The difference between normal and abnormal heat conduction behaviors can be further appreciated also in the steady-state kinetic temperature profiles $T(x)$. For systems with normal heat conduction $T(x)$  is determined by solving the stationary heat equation assuming that the thermal conductivity is proportional to $\sqrt{T}$ as prescribed by standard kinetic theory, yielding~\cite{A. Dhar2001}
\begin{equation}\label{Eq4}
T(x)=\left[T_{h}^{3/2}\left(1-\frac{x}{L}\right)+T_{c}^{3/2}\frac{x}{L}\right]^{2/3}.
\end{equation}
In Fig.~\ref{fig3}(a) this prediction is compared with our simulation results for $W=1$. It is seen that there is a good agreement between the results of our numerical simulations and
Eq.~(\ref{Eq4}). To better appreciate the deviations from the prediction, we also plot the differences $\delta T$ between the data and the black line. It is shown in the inset of Fig.~\ref{fig3}(a) that $|\delta T|$ decreases with increasing $L$, as expected since $|\nabla T|=\Delta T/L$ decreases when $L$ increases, indicating that the linear response can correctly describe the transport properties of the system for large enough system length. On the other hand, for systems with abnormal heat conduction the temperature profile is expected to be qualitatively different, being solution of a fractional diffusion equation as demonstrated in several cases \cite{S. Lepri2011,Kundu2019}. A typical feature is that the temperature profile is concave upwards in part of the system and concave downwards elsewhere, and this is true even for small temperature differences~\cite{T. Mai2007,A. Lippi2000,S. Lepri2009}. This is confirmed in Fig.~\ref{fig3}(b) by our numerical simulations for $W=0.1$. Note that the data for three different $L$ overlap with each other, implying that the deviations from Fourier's behavior are not finite-size effects. Altogether, those numerical results again support our findings based on the length-dependence of the thermal conductivity, that heat conduction in 3D is normal while in lower dimensions it is abnormal.\par

We now turn to the comparison with linear-response results. Based on the celebrated Green-Kubo formula, which relates transport coefficients to the current time-correlation functions, the thermal conductivity can be expressed as~\cite{R. Kubo1991,S. Lepri2003,A. Dhar2008}
\begin{equation}\label{Eq5}
\kappa_{GK}=\frac{\rho}{k_BT^2}\lim_{\tau_{\mathrm{tr}}\rightarrow\infty}\lim_{N\rightarrow\infty}\frac{1}{N}\int_{0}^{\tau_{\mathrm{tr}}}\left<J(0)J(t)\right>dt.
\end{equation}
In this formula, $J\equiv\frac{1}{2} \Sigma_{i}^{N}\mathbf{v}_{i}^2v_{x,i}$ represents the total heat current along the $x$ coordinate and $\left<J(0)J(t)\right>$ is its correlation function  in the equilibrium state.  In the simulations, we consider an isolated fluid with periodic boundary conditions also in the $x$ direction. The initial condition is randomly assigned with the constraints that the total momentum is zero and the total energy corresponds to $T=1$. The system is then evolved and after the equilibrium state is attained, we compute $\left<J(0)J(t)\right>/N$ and the integral in~Eq.~(\ref{Eq5}).\par

\begin{figure}
\includegraphics[width=10cm]{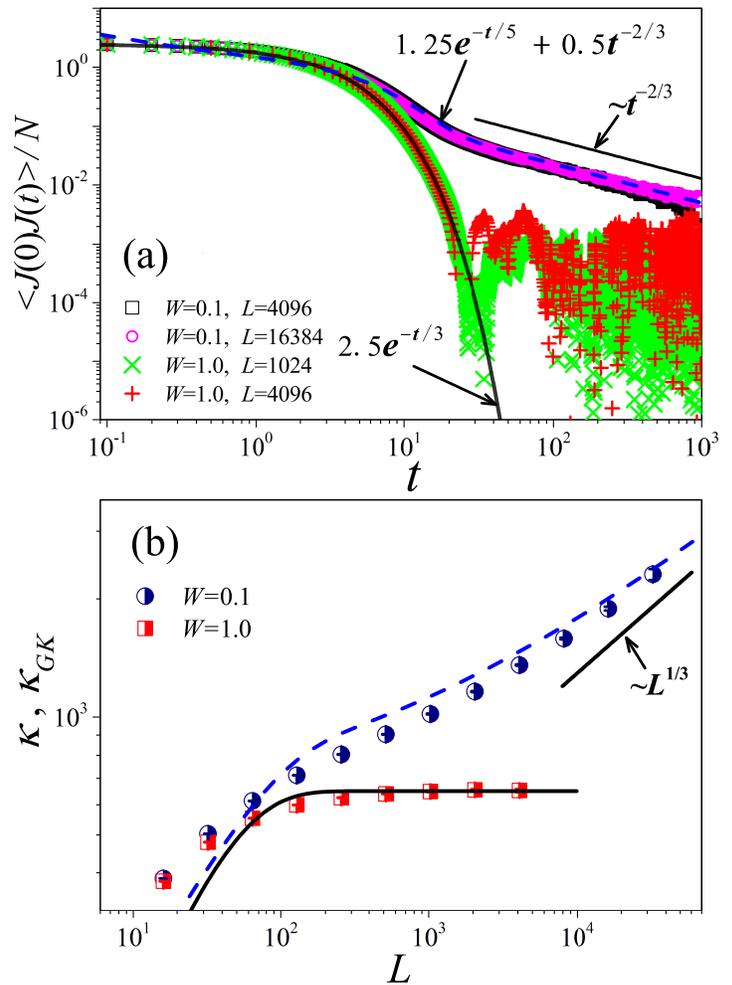}
\caption{(Color online) (a) Correlation functions of the total heat current for the 3D momentum-conserving fluid system. The black solid line and the blue dashed line are the fit of the data with $W=1.0$ and $W=0.1$, respectively. (b) the comparison of the thermal conductivity obtained by using the Green-Kubo formula [Eq.~(\ref{Eq5})] and by using the nonequilibrium setting (symbols). The integration is up a time $L/C$ with $C=12.0$ fixed empirically (see text). }
\label{fig4}
\end{figure}

The results for different $W$ values are presented in Fig.~\ref{fig4}. It can be seen from Fig.~\ref{fig4}(a) that for $W=1$, the correlation function undergoes a rapid decay at short times, and eventually, when $t>20$, it begins to oscillate around zero, (the negative values of $\left<J(0)J(t)\right>/N$ are not shown in this log-log scale.).
Fitting exhibits an exponential decay. However, it is clear that for $W=0.1$, after a rapid decay at short times, the correlation function eventually follows a power law decay of $\left<J(0)J(t)\right>/N\sim t^{\gamma}$, fully compatible with the theoretical prediction $\gamma=-2/3$.

As a further check, in Fig.~\ref{fig4}(b), we compare the nonequilibrium values of $\kappa$ (symbols) with the finite-size Green-Kubo value $\kappa_{GK}(L)$ (lines). The latter is obtained truncating the integral in Eq.(\ref{Eq5}) up to $\tau_{\mathrm{tr}}=L/C$ where $C$ is a constant
that we fix empirically to yield the best match between $\kappa$ and
$\kappa_{GK}(L)$. This procedure is justified by the fact that for finite systems the integral in Eq.(\ref{Eq5}) should be extended up to times of order of the characteristic transit time of waves $L/c_{s}$ ($c_{s}$ being the sound speed)~\cite{S. Lepri2003,A. Dhar2008}. It can be seen that $\kappa_{GK}(L)$ agrees with $\kappa$, especially the part for large system length. Comparing the results for $W=1$ and $W=0.1$, we conclude that for a fixed $L$ (see $L=4096$), a gradual transition from 3D to 1D behavior of the thermal conductivity will take place when $W$ is decreased. Note also that, both in the 3D and quasi-1D cases the conductivity should approach the bulk value for $L$ larger than the mean-free-path which the same in both cases. So it is reasonable that the two curves start to differ from a size $L\approx 100$ much larger the mean free path for $\tau=0.1$. Once again, the emerging picture supports the validity of the Fourier law for large $W$ and the dimensionality-crossover behavior for small $W$.\par

\textit{\textbf{Summary and discussion.}} We have numerically studied heat conduction in a 3D fluid by nonequilibrium thermal-wall and equilibrium Green-Kubo methods. Both  give sound evidence that in the 3D case i.e. for $W$ is large enough, the system has a finite nondiverging thermal conductivity. As $W$ decreases, the system exhibits a crossover from 3D to 1D behavior of the thermal conductivity. These results are rooted in the condition for 3D fluid systems described by MPC dynamics to keep the momentum conservation of the system. The fact that the microscopic MPC dynamics is inherently stochastic does
not alter the scenario,  as seen by comparing with anharmonic lattices with symplectic dynamics~\cite{K. Saito2010}. Also we provided a strong evidence that the Fourier law is valid in 3D fluid systems, further confirming that the predictions of existing theories~\cite{S. Lepri2016,G. Basile2006} apply to fluids. In addition, our results suggest that the dimensionality-crossover behavior of heat conduction in 3D lattices~\cite{K. Saito2010,L. Wang2010} can also take place in 3D fluids. Remarkably, the quasi-1D fluid displays the same transition from a normal to abnormal behavior for large enough longitudinal sizes, characteristic of strictly 1D models \citep{Zhao2018,Miron2019,S. Lepri2020,Lepri2021}. This stems from the fact that abnormal (hydrodynamic) effects can be only observed on scales that can be much larger than the typical kinetic mean-free-path \citep{S. Lepri2020}. Apart from general theoretical implications in transport theory, our findings may have experimental relevance as well, because in experimental setup researchers usually fix the cross-sectional area of samples and study how the thermal conductivity changes with the sample's length for achieving the purpose of understanding and controlling thermal transport. \par

\textit{\textbf{Acknowledgments.}} We acknowledge support by the Education Department of Fujian Province (Grant No. JAT200035).\par

\end{document}